\documentclass[prb,twocolumn,showpacs]{revtex4}

\usepackage{graphicx}
\usepackage{dcolumn}
\usepackage{amsmath}
\usepackage{color}

\begin{document}
\title{Korringa-like relaxation in the high-temperature phase of $A$-site ordered YBaMn$_2$O$_6$}

\author{S.~Schaile}
\author{H.-A.~Krug von Nidda}
\author{J.~Deisenhofer}
\author{A.~Loidl}
\affiliation{Experimentalphysik V, Center for Electronic
Correlations and Magnetism, Institute for Physics, Augsburg
University, D-86135 Augsburg, Germany}

\author{T.~Nakajima}
\author{Y.~Ueda}
\affiliation{Material Design and Characterization Laboratory, Institute for Solid State Physics, University of Tokyo, 5-1-5 Kashiwanoha, Kashiwa, Chiba 277-8581, Japan}

\date{\today}

\begin{abstract}
We report on high-temperature electron spin resonance studies of
$A$-site ordered YBaMn$_2$O$_6$ and disordered
Y$_{0.5}$Ba$_{0.5}$MnO$_3$. In the disordered sample we find that
the linewidth is governed by spin-spin relaxation processes as in
many other manganite systems. In contrast we find a Korringa-like
spin relaxation with a slope of about 1 Oe/K above the
charge-ordering transition and extending up to 930~K in the metal
ordered YBaMn$_2$O$_6$ samples. A Korringa-law is a clear feature of
a truly metallic state, pointing to an unconventional
high-temperature phase of these ordered manganites. In agreement
with the ESR intensity this suggests that in the metallic
high-temperature phase the ESR signal stems from Mn$^{4+}$ core
spins which relax via the quasi-delocalized $e_g$ electrons.

\end{abstract}



\maketitle
\section{Introduction}

During the last decades manganese-based oxides have been in the
focus of condensed matter research due to their rich phenomenology
ranging from colossal magnetoresistance effects to charge/orbital
ordering and to multiferroicity. All of these aspects are predicted to occur in the phase space of half-doped
manganites with chemical formula $A_{0.5}A^\prime_{0.5}$MnO$_3$ with
$A$ a trivalent rare-earth ion and $A^\prime$=
(Ca,Sr,Ba). The role of disorder due to a random distribution of $A$
and $A^\prime$ ions was found to be a decisive ingredient for this
class of materials and, in particular, the possibility of growing
both disordered $A_{0.5}$Ba$_{0.5}$MnO$_3$ and ordered
$A$BaMn$_{2}$O$_6$ systems revealed significant differences in the
corresponding phase diagrams:\cite{Nakajima02,Akahoshi03,Nakajima05}

The $A$-site ordered manganites with perovskite structure contain
two different ions ($A$ and $A'$). These ions are arranged in two
alternating $A$O and $A'$O layers which are stacked along the
$c$-axis, separated by MnO$_6$ octahedra. $A$-site order as high as
96\% can be reached.\cite{Nakajima04} With exchanging the rare earth
ions on the $A$-site and Ba ions residing on the $A'$-site a rich
phase diagram can be obtained. With increasing ionic radius on the
$A$-site the system evolves from an orbital-ordered/charge-ordered,
antiferromagnetic ground state to a ferromagnetic ground state for
large ions.\cite{Akahoshi03} The different valence of trivalent rare-earth ion and divalent Ba ion is leading to mixed valence on the
Mn sites, resulting in a mixture of Mn$^{3+}$ and Mn$^{4+}$ ions.
YBaMn$_2$O$_6$, with the largest discrepancy of the ionic radii
between $A$ and $A'$-site, exhibits three peaks in the specific heat
at 200\,K ($T_{c3}$), 480\,K ($T_{c2}$) and 520\,K ($T_{c1}$).
Conductivity measurements show a metal-to-insulator transition at
$T_{c2}$ \cite{Nakajima02} and neutron-diffraction studies reveal a
structural phase transition from monoclinic above $T_{c1}$ to
triclinic below. \cite{Williams05} The insulating phase below
$T_{c2}$ has been associated with charge and orbital order,
accompanied by antiferromagnetic ordering below $T_{c3}$.
\cite{Daoud-Aladine08} The two-step charge-ordering transition is
accompanied by a strong change of the effective magnetic moment from
$\mu_{eff}=10.35\mu_B(=7.32\mu_B$ per Mn-ion) below $T_{c2}$ to
$\mu_{eff}=6.22\mu_B(=4.40\mu_B$ per Mn-ion) above $T_{c1}$. The
susceptibility follows a Curie-Weiss law with $\Theta=-512$\,K below
the charge-ordering transition and $\Theta=398$\,K above. In
contrast, the disordered system Y$_{0.5}$Ba$_{0.5}$MnO$_3$ does not
show any charge- and orbital ordering, but exhibits spin-glass like
features below 30\,K.\cite{Nakajima04}

Electron spin resonance (ESR) provides  a local probe to study the
degrees of freedom of electron, spin, charge and orbitals and their
coupling to the lattice
\cite{Huber99,Ivanshin00,Kochelaev03,Deisenhofer05} which are
considered to be the origin of the complex ordering phenomena in
manganites. In our previous study on YBaMn$_2$O$_6$ we reported ESR
data up to a temperature of 600 K.\cite{Zakharov08} We show that the
behavior of all ESR parameters in the low-temperature charge-ordered
regime is strongly distinguished from that in the high-temperature
metallic regime and with jump-like changes at $T_{c1}$. Focusing on
the temperature dependence of the linewidth, the charge-ordered
phase is strongly reminiscent of orbital-ordered
La$_{1-x}$Sr$_x$MnO$_3$ ($x<0.15$), where the linewidth follows the
temperature dependence of the orbital-order
parameter.\cite{Ivanshin00,Deisenhofer03,Alejandro03} In contrast,
at high temperatures a quasi-linear increase suggests a
Korringa-type relaxation. However, due to the limited temperature
range above $T_{c1}$ a thermally activated behavior as reported for
La$_{1-x}$Ca$_x$MnO$_3$\cite{Shengelaya00} or a spin-spin relaxation
dominated mechanism\cite{Huber99} could not be excluded,
previously. In contrast to the static susceptibility the effective
moment derived from the ESR intensity as
$\mu_{ESR}=5.0(3)\mu_B(=3.53\mu_B$ per Mn-ion) is within the
expected range of the magnetic moment of Mn$^{4+}$ ions only,
suggesting that the $e_g$ electrons of the Mn$^{3+}$  do not
contribute to the resonance absorption due to fast hopping between
the Mn$^{4+}$ cores.\cite{Zakharov08}

Here we report on measurements up to 930\,K, which
unambiguously establish a Korringa-like increase and, hence,
identify a truly metallic feature in this $A$-site ordered
manganite. This behavior is contrasted with the results obtained for
disordered Y$_{0.5}$Ba$_{0.5}$MnO$_3$.

\section{Experimental details}

Polycrystalline samples of the ordered system were prepared by
solid-state reaction of R$_2$O$_3$, BaCO$_3$, and MnO$_2$ in Ar flow
at 1573\,K for 48\,h. After sintering the oxygen deficiency in
YBaMn$_2$O$_{6-y}$ was about $y\sim 0.9$, but a final oxygen content
of 6.00(2) could be realized by post-annealing at 773K in pure
oxygen for 48\,h. Annealing at higher temperatures results in an
exchange of $A$ and $A`$-site ions leading to disorder in the YO and
BaO planes.\cite{Nakajima04} The ordering of the Y$^{3+}$ and the
Ba$^{2+}$ cations was checked by powder x-ray diffraction (occupancy
refinements) resulting in an ordering degree of nearly 100\%. The
disordered sample was prepared by the same starting materials and
also showed full oxygen occupancy. Mixed powder was sintered at 1623
K in 1\% O$_2$/Ar flow for 48\,h and then heated at 1173\,K in O$_2$
flow for 48\,h. The degree of disorder has been evaluated by
Rietveld refinements similar to the procedure described in Ref.~4. Within experimental uncertainty of the powder XRD, no sizeable impurities could be determined for both systems.

For ESR measurements in the temperature region $300 < T < 1000$\,K a
Bruker Elexsys II spectrometer equipped with a Bruker ER 4114 HT
X-Band ($\nu$=9.49\,GHz) cavity was used. ESR detects the power $P$
absorbed by the sample from the transverse magnetic microwave field
as a function of the static magnetic field $H$. The signal-to-noise
ratio of the spectra is improved by recording the derivative $dP/dH$
using lock-in technique with field modulation. Due to large
temperature gradients at high temperatures the sample temperature is
corrected by calibration measurements at the actual sample site in
the high temperature cavity. To check the influence of different
environmental conditions and temperature on the oxygen content of
the samples at high temperatures, measurements have been performed
in pure Ar at 0.7 bar pressure adjusted at room temperature and in air
(open quartz tube).

\begin{figure}[h]
\includegraphics[width=0.9\columnwidth]{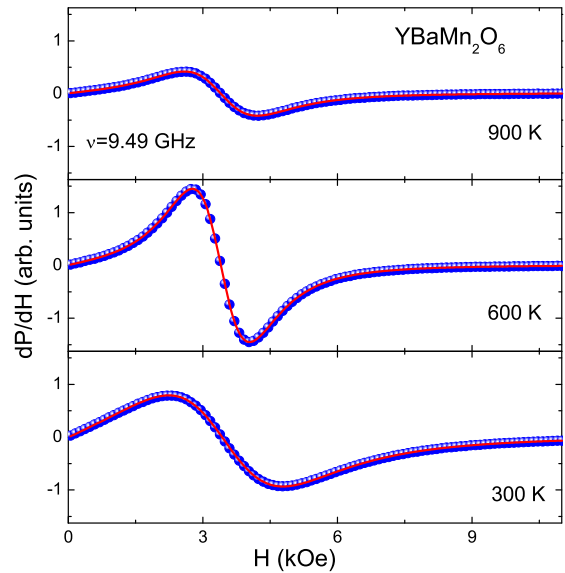} \caption{\label{fig:spek} ESR spectra of YBaMn$_2$O$_6$ at different temperatures below
 and above the charge-ordering transition ($T_{c2}$=480\,K). The red line refers to the corresponding fit curves using Lorentzian line shapes (first derivative). The number of data points is reduced by a factor of 10.}
\end{figure}

\section{Experimental Results and Discussion}
Fig.~\ref{fig:spek} shows ESR spectra of ordered YBaMn$_2$O$_6$ for
different temperatures in the paramagnetic regime. Even at elevated
temperatures the signal is nicely detectable. The signal-to-noise
ratio remains large up to highest temperatures, thus the absorption
can be evaluated with high accuracy and can be described well by a
single exchange narrowed Lorentzian line in the whole temperature
range. Due to the large linewidth the resonance at $-H_{res}$ has to
be included.\cite{Joshi04} In a usual metal one would expect a
Dysonian line shape,  but the degree of admixture of dispersion
depends on the actual grain size and skin depth. The resistivity
value reported for this polycrystalline material\cite{Zakharov08} is high compared to
a good metal, leading to a skin depth larger than the grain size and
a Lorentzian line shape.\cite{Ivanshin00}

As expected for transition metals with less than half filled
\textit{d} shells,  the effective $g$-factor remains constant in the
metallic high-temperature regime at a value of $g=1.985$ (Fig. 2,
lower frame) which within experimental uncertainty corresponds to
the insulator single-ion value of Mn$^{4+}$ $g=1.994$ found in cubic
symmetry.\cite{Abragam70} Below $T_{c1}$ the $g$-factor slightly
drops because of charge fluctuations and charge-ordering processes.
The ESR intensity follows a Curie-Weiss law with $\Theta=387$\,K in
the high-temperature phase, indicating strong ferromagnetic
correlations (Fig. 2, middle frame). Considering an uncertainty of
about 20\,K in $\Theta$ due to uncertainties in the quality factor
of the cavity and sample temperature in the high temperature region
this value fits nicely to the values derived from susceptibility
measurements. \cite{Nakajima02}

\begin{figure}[t]
\includegraphics[width=0.9\columnwidth]{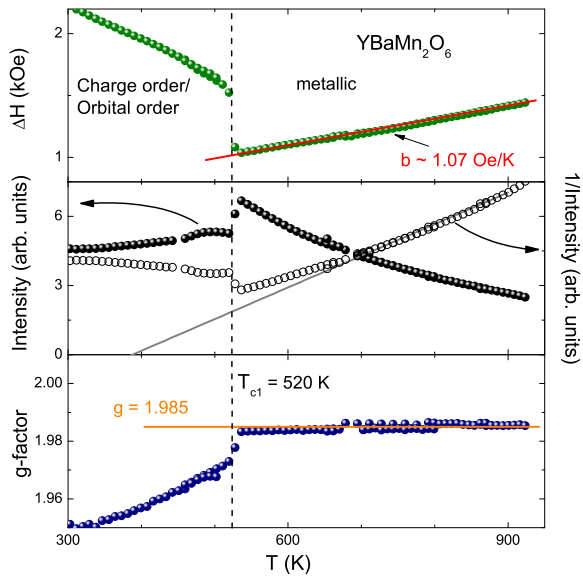}
\caption{\label{fig:dH} Upper frame: Temperature dependence
of the ESR linewidth $\Delta H$ in $A$-site ordered YBaMn$_2$O$_6$. A Korringa-type fit is indicated by a solid line.
 Middle frame: Temperature dependence of the ESR intensity and the inverse of the intensity. A Curie-Weiss fit is indicated by a solid line.
  Lower frame: Temperature dependence of the effective $g$ factor. A constant $g$ value $g = 1.985$ is indicated by a solid line. The vertical dashed line is indicating the temperature of the charge-ordering transition ($T_{c1}$=520\,K).}
\end{figure}

The temperature dependence of the ESR linewidth $\Delta H$ from
$300$\,K to $930$\,K is shown in the upper frame of Fig.~\ref{fig:dH}. As
reported previously, in the low-temperature regime, between $T_{c3}
= 200$\,K and $T_{c1} = 520$\,K, the ESR linewidth decreases from
3\,kOe to 1\,kOe with increasing temperature.\cite{Zakharov08} Our
new data reveal a sharp drop from 1.5\,kOe to  1\,kOe at the
 structural phase transition at $T_{c1} = 520$\,K in agreement with a
 first-order transition. Above the structural transition at $T_{c1} = 520$\,K the linewidth
increases with increasing temperature and shows a linear behavior
which can be described by
\begin{equation} \label{eqn:korringa}
\Delta H = \Delta H_0 + b T
\end{equation}
with a zero-temperature value $\Delta H_0 = 420$\,G and slope
$b=1.07$\,Oe/K. This value for the slope $b$ differs by a factor of
about two from the one obtained previously by fitting in a narrow
temperature range above $T_{c1}$. This discrepancy stems from a
temperature gradient between the sample's position inside the quartz
tube and the position of the temperature probe. This gradient turned out to become significant above 500\,K which is given as specification limit of the Bruker ER 4141VT System used previously.
As mentioned above our new data has been obtained using an in-situ calibration of the
temperature at the sample position, thus ensuring a high accuracy of
the actual sample temperature. All parameters coincided for heating and cooling.

\begin{figure}[t]
\includegraphics[width=0.9\columnwidth]{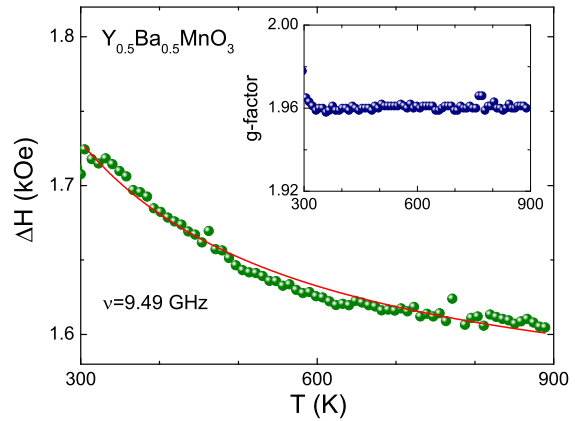}
\caption{\label{fig:disorder} Temperature dependence of the ESR linewidth
 $\Delta H$ in $A$-site disordered Y$_{0.5}$Ba$_{0.5}$MnO$_3$ together with fit following Eqn.\eqref{eqn:Kubo}.
 Inset: Temperature dependence of the effective $g$ factor.}
\end{figure}

To observe such a linear temperature dependence over a broad
frequency range is rather uncommon for manganites, where the
linewidth usually tends to saturate at high temperatures. This
standard behavior is illustrated in Fig.~\ref{fig:disorder} for the
related disordered system Y$_{0.5}$Ba$_{0.5}$MnO$_3$. With
increasing temperature the linewidth monotonously decreases
approaching an asymptotic high-temperature value of about
$1.5$\,kOe. At the same time the $g$ factor is found to be
temperature independent at $g \approx 1.96$. Note that this value is
close to that of YBaMn$_2$O$_6$ in the charge-ordered phase. The
linewidth can be well described by the Kubo-Tomita approach for
spin-spin relaxation in insulating paramagnets as applied in
manganites by Huber $et~al$.\cite{Huber99}
\begin{equation} \label{eqn:Kubo}
\Delta H(T) = \frac{\chi_0(T)}{\chi_{\rm dc}(T)}\Delta H_{\infty} = \frac{T-\Theta}{T}\Delta H_{\infty},
\end{equation}
where $\chi_0(T)$ and $\chi_{\rm dc}(T)$ describe the
susceptibilities of free and interacting spins, respectively, and
$\Delta H_{\infty}$ denotes the asymptotic high-temperature limit of
the linewidth. The fit yields a Curie-Weiss temperature $\Theta=-38$\,K, in reasonable
agreement with the susceptibility ($\Theta=-18$\,K determined from SQUID measurements), and $\Delta H_{\infty} =
1.5$\,kOe which is comparable to the values in many other manganite
systems where the relaxation was found to be dominated by pure
spin-spin relaxation
processes.\cite{Huber99,Ivanshin00,deisenhofer02}

Three types of relaxation mechanisms are known to result in a
strictly linear temperature dependence of the linewidth. The first
corresponds to a modulation of the crystalline electric field by
lattice vibrations, where only one phonon is involved in the
relaxation process.\cite{Abragam70} In the case of dilute magnetic
ions in a non-magnetic host lattice, this \textit{direct spin-phonon
process} yields a relaxation rate
\begin{equation} \label{eqn:directprocess}
\Delta H \propto \omega^n \coth\left(\frac{\hbar \omega}{2k_B T}\right)
\end{equation}
with $\omega$ being the resonance frequency, and $n=3$ or 5
corresponding to non-Kramers or Kramers ground states of the
magnetic ions, respectively. In the limit $k_B T\gg \hbar \omega$ the linewidth
can be approximated by
\begin{equation} \label{eqn:directprocess2}
\Delta H \propto \omega^{n-1} T.
\end{equation}
The strong dependence on the resonance frequency or magnetic field
is, however, not necessarily a characteristic property anymore in
concentrated exchange-coupled magnetic systems.\cite{Seehra68} A
linear increase stemming from a direct phonon process was suggested
for CrBr$_3$, NiCl$_2$ and other systems where the magnetic ion has
no half-filled shell and spin $S\geq 1$.\cite{Birgeneau73,Huber75}

The second scenario was proposed by Seehra and Castner, where a
linear temperature dependence of the linewidth can dominate the line
broadening at temperatures of the order of the Debye temperature, if
a large enough \textit{static Dzyaloshinsky-Moriya interaction}
$\mathbf{D}(\mathbf{S_i}\times \mathbf{S_j})$ between a pair of
magnetic ions with spin $S_i$ and $S_j$ exists and is modulated by a
single phonon for temperatures much higher than the Zeeman splitting
$k_BT\gg g\mu_B H$ and exchange coupling constants $J\gg g\mu_B
H$:\cite{Seehra68,Castner71}
\begin{equation} \label{eqn:modulatedDM}
\Delta H \simeq \frac{4}{9}\frac{z}{g \mu_B}\frac{(\lambda R)^2 |\mathbf{D}|^2J^2}{\rho\hbar^2}
\left\langle \frac{1}{c_t^5}+\frac{2}{3}\frac{1}{c_l^5}\right\rangle k_B T
\end{equation}
Here $z$ is the magnetic coordination number, $\rho$ the crystal
density, $c_{t,l}$ the transverse and longitudinal sound
velocity, and $R$ the nearest-neighbor distance.
The parameter $\lambda J \sim dJ/dr$ parameterizes the modulation of the exchange constant $J$ by the ionic
displacements due to the phonon and is a measure of the spin-phonon
coupling.\cite{Baltensperger68,Kant10,Kant12} Such a scenario has
been proposed to be realized in the spin $S=1/2$ systems
Cu(HCOO)$_2\cdot$ 4 H$_2$O \cite{Seehra68} and
SrCu$_2$(BO$_3$)$_2$,\cite{Zorko04} but is usually overruled by the influence of the crystal electric field for systems with larger spins.

In early ESR studies in manganites the former \textit{direct phonon
process} had been evoked to describe the quasi-linear behavior of
the linewidth in some disordered
manganites.\cite{Seehra96,Lofland97,Rettori97} However, it was shown
subsequently that a saturation behavior is reached at higher
temperatures, which is characteristic of dominating spin-spin
relaxation processes as described above for disordered
Y$_{0.5}$Ba$_{0.5}$MnO$_3$.\cite{Huber99} Therefore, we discard the
direct-phonon scenario for the high-temperature phase of
YBaMn$_2$O$_6$ and will concentrate on the third possibility to
describe a linear temperature dependence, namely the well-known
\textit{Korringa-type relaxation}, which in contrast to the above
mechanisms is governed by the presence of itinerant charge carriers
in agreement with the metallic features of the high-temperature
state, in particular the observation that the $e_g$-electrons seem
to act as quasi-delocalized and do not participate in the resonant
phenomenon as reported previously.\cite{Zakharov08}

The Korringa-law for the linewidth is given by
\begin{equation}\label{Eq:Korringa}
\Delta H \propto \langle J_{\rm CE-ls}^2(q) \rangle N^2(E_{\rm F}) T,
\end{equation}
where the slope is proportional to the square of the electronic
density of states $N(E_{\rm F})$ at the Fermi energy $E_{\rm F}$ and
the exchange coupling $J_{\rm CE-ls}$ between the conduction electrons
and the localized spins averaged over the momentum transfer $q$ of the conduction electrons. The rather low value of $b\simeq 1$
Oe/K might be indicative of a low-density of state at the Fermi
energy, but can also originate from the peculiarity of the
manganite systems in terms of a bottlenecked relaxation as discussed
below. The quantity $\Delta H_0$ is affected by the relaxation times
of ions and electrons, but is also strongly affected by
crystal-field contributions, dipole-dipole interactions or lattice
defects.\cite{Barnes81} It is known, that these secondary effects
\cite{Kochelaev94} overlay the intrinsic contributions so strongly
that a quantitative evaluation of this parameter is not possible. In
contrast the Korringa-slope is robust with respect to secondary
contributions.

Because of a lack of reports on Korringa-type relaxation in
compounds containing Mn$^{3+}$ or Mn$^{4+}$ we have to compare the
Korringa-slope to Mn$^{2+}$ systems with half-filled $d$-orbitals.
This scenario is closer to the situation of the half filled $t_{2g}$
orbitals in Mn$^{4+}$ than to partially filled $e_g$ orbitals in
Jahn-Teller active Mn$^{3+}$. For Mn$^{2+}$ a bottlenecked
Korringa-relaxation is the usual case, which is also of importance
in manganese doped La$_{2-x}$Sr$_x$CuO$_4$.\cite{Kochelaev94} This
bottleneck can be opened by doping so that for example a Korringa
slope of 45\,Oe/K can be seen in unbottlenecked relaxation of highly
diluted Mn$^{2+}$ in Ag:Mn:Ni.\cite{Loidl97} Other ESR results also
support a bottlenecked spin relaxation in perovskite
manganites.\cite{Shengelaya00,Padmanabhan07}

\begin{figure}[t]
\includegraphics[width=0.8\columnwidth]{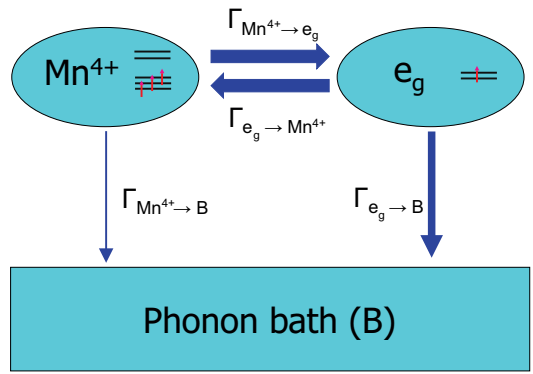}
\caption{\label{fig:Bottleneck} Schematic sketch of the
bottleneck scenario for the spin relaxation of the local Mn$^{4+}$ moments via the highly mobile $e_g$ electron system in the metallic regime. The thickness of the arrows
indicates the effectiveness of the relaxation paths.}
\end{figure}

In the case of YBaMn$_2$O$_6$ a strong bottleneck scenario seems
applicable, too. This bottleneck can be found in the relaxation of
the conduction electrons to the phonon bath (B) $\Gamma_{e_g \rightarrow
B}$, which is slower than the fast Overhauser relaxation
$\Gamma_{e_g \rightarrow Mn^{4+}}$ which supports a backscattering
of the excitation from the conduction electrons to the Mn$^{4+}$
spin before it can transfer its energy to the phonon bath. Thus the Korringa-slope is
reduced by the ratio:
\begin{equation}
\frac{\Gamma_{e_g \rightarrow B}}{\Gamma_{e_g \rightarrow Mn^{4+}}+\Gamma_{e_g \rightarrow B}}
\end{equation}
The relaxation path between Mn$^{4+}$ spins and phonon bath is
negligible  because of weak $S$-state interactions whereas the cross
relaxation between Mn$^{4+}$ spins and the conduction electrons is
triggered by strong Hund's coupling and, thus, dominates the
relaxation of the conduction electrons to the phonon bath even at
highest temperatures. The linear temperature dependence of the
linewidth is not affected, because the cross relaxation between
Mn$^{4+}$ spins and conduction electrons is temperature
independent.\cite{Barnes81} The relaxation of the conduction
electrons to the phonon bath can be anticipated to be temperature
independent as well, because the conductivity changes only weakly
with temperature above $T_{c1}$.\cite{Nakajima04} The bottleneck scenario is further supported by the fact, that we do not observe any significant shift of the $g$-value which in contrary would be expected in the case of non-bottlenecked Korringa relaxation.

At present, there does not seem to exist any microscopic model for
mixed-valence manganites, which predicts a Korringa-like
spin-relaxation rate as observed. However, Huber \textit{et al.}
suggested that mixed-valence manganites may be separated into a
$S=3/2$ system originating from the Mn$^{4+}$ $t_{2g}$ core spins
and a quasi itinerant $S=1/2$ system corresponding to the $e_g$
electrons,\cite{Huber07} which would support our bottleneck
scenario. Moreover, Moskvin proposed that the high-temperature,
orbitally disordered state of LaMnO$_3$ may be described by an
electron-hole Bose liquid, which results from a charge-transfer
instability leading to effective Mn$^{2+}$-Mn$^{4+}$ configurations
and metallic-like behavior.\cite{Moskvin09} Similarly, such a
mechanism may be realized in our system above the charge- and
orbital-ordering transition. One may speculate that the
Korringa-like behavior of the ESR linewidth could be a signature of
such a charge-disproportion picture and we hope that our results
will foster further theoretical efforts with regard to
spin-relaxation in manganites.

\section{Conclusions}
In summary, we investigated the high-temperature ESR behavior in
polycrystalline samples of ordered YBaMn$_2$O$_6$ and disordered
Y$_{0.5}$Ba$_{0.5}$MnO$_3$. The latter exhibits a conventional
temperature dependence of the linewidth due to spin-spin relaxation.
The linear high-temperature increase of the linewidth over a
temperature range of 400\,K is a unique feature for $A$-site ordered
YBaMn$_2$O$_6$. It can be well described by a bottlenecked Korringa
relaxation and indicates a scenario where highly mobile $e_g$
electrons are present on a Mn$^{4+}$ background. This is in accordance
to the observed magnetic moment for Mn$^{4+}$ only, obtained from
the ESR intensity. The low $g$-shift matches the nearly quenched orbital moment of the
half-filled $t_{2g}$ shell of this Mn$^{4+}$ background as well as the bottleneck scenario.
To further analyze this mechanism single crystal measurements should be
performed to enhance the accuracy of conductivity measurements at
high temperatures and give an insight into a possible anisotropy of
the spin-relaxation.

\begin{acknowledgments}
It is a pleasure to thank D. L. Huber, M. V. Eremin, A. S. Moskvin, and
N. Pascher for fruitful discussions. We are grateful to Dana Vieweg
for susceptibility measurements. This work is partially supported by
the Deutsche Forschungs Gemeinschaft via the Collaborative Research
Center TRR 80 (Augsburg-Munich).
\end{acknowledgments}


\end{document}